# OOD-Speech: A Large Bengali Speech Recognition Dataset for Out-of-Distribution Benchmarking

*Fazle Rabbi Rakib* [*,1,2], *Souhardya Saha Dip* [*,1,2], *Samiul Alam* [1], *Nazia Tasnim* [1], *Md. Istiak Hossain Shihab* [1],
*Md. Nazmuddoha Ansary* [1], *Syed Mobassir Hossen* [1], *Marsia Haque Meghla* [1], *Mamunur Mamun* [1], *Farig Sadeque* [1,3],
*Sayma Sultana Chowdhury* [2], *Tahsin Reasat* [1,4], *Asif Sushmit* [†,1,5], *Ahmed Imtiaz Humayun* [†,1,6]

[1]Bengali.AI, [2]SUST, [3]Brac University, [4]Vanderbilt University, [5]RPI, [6]Rice University

{farig,reasat,sushmit,imtiaz}@bengali.ai

## Abstract

We present OOD-Speech, the first out-of-distribution (OOD) benchmarking dataset for Bengali automatic speech recognition (ASR). Being one of the most spoken languages globally, Bengali portrays large diversity in dialects and prosodic features, which demands ASR frameworks to be robust towards distribution shifts. For example, islamic religious sermons in Bengali are delivered with a tonality that is significantly different from regular speech. Our training dataset is collected via massively online crowdsourcing campaigns which resulted in 1177.94 hours collected and curated from 22,645 native Bengali speakers from South Asia. Our test dataset comprises 23.03 hours of speech collected and manually annotated from 17 different sources, e.g., Bengali TV drama, Audiobook, Talk show, Online class, and Islamic sermons to name a few. *OOD-Speech is jointly the largest publicly available speech dataset, as well as the first out-of-distribution ASR benchmarking dataset for Bengali.*

## 1. Introduction

The field of speech analysis has undergone widespread development with the rise in popularity of virtual platforms for communication and learning. Apart from major contributions in technology accessibility [1], Text-to-Speech (TTS) and Automatic Speech Recognition (ASR) have ubiquitous applications across the spectrum, e.g., Automatic Speech Assessment for language education [2], and language disorder assessment and therapy [3]. In all such applications, the diversity of training datasets play a crucial role. For instance, if a speech recognition system is trained on only one English accent, the trained model would be susceptible to failure and be less robust for other dialects and variants of English. This is termed the Out-of-Distribution (OOD) effect, and is an active area of study in Computer Vision with a number of standard benchmarking datasets and competitions [4]. Contrarily, datasets for OOD benchmarking are not widespread in ASR, with researchers mostly relying on multiple datasets and assuming they are OOD [5].

The diverse dialects/accents and unique morphology of the Bengali language (Suppl. Sec.B) makes the requirement of large datasets even more imperative for robust training. Over the years, a number of datasets have been assembled for Bengali, each having its own limitations regarding speaker diversity, size, and public availability [8, 9, 10]. We provide comparisons in Suppl.

Symbols * and † denote equal contribution.

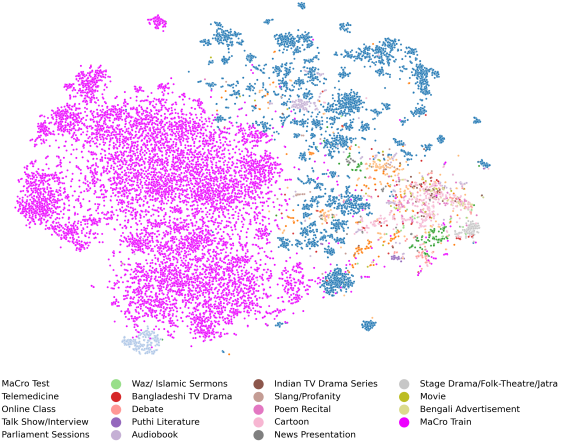

Figure 1: *t-Stochastic Neighbor Embeddings [6] of Geneva features [7] extracted from MaCro Train, MaCro Test and OOD Test subsets from our dataset. We can see a clear separation between the clusters formed by MaCro Train and OOD Test. This indicates that the subsets are clearly different in the Geneva feature space.*

Tab. 3. To facilitate research in this direction, we present *OOD-Speech*, a large publicly available Bengali dataset for out-of-distribution benchmarking. Our dataset contains data collected via two major collection thrusts. The first thrust incorporated large scale online crowdsourcing campaigns to collect scripted speech from various accents and uncontrolled environments(Suppl. Sec.C). This resulted in our MaCro or **Ma**ssively **Cro**wdsourced speech subset, comprising 1188.14 hours of speech data collected from 22,645 contributors from India and Bangladesh. The second thrust was to collect and manually annotate spontaneous speech from sources available on the internet. This resulted in the OOD Subset or **Out-of-D**istribution spontaneous speech subset, containing 13.19 hours of speech data collected from 2768 YouTube videos and miscellaneous sources. The training set of OOD-Speech contains 1129.46 hours worth of audio recordings of 934084 Bengali sentences sampled from Wikipedia or contributed by volunteers. The test set contains two subsets of data, 9.84 hours from MaCro and 13.19 hours from the OOD subset, which comprised natural speech from 17 different domains available on the internet. Ours is the largest publicly available ASR dataset for Bengali(Suppl. Tab.3), and also the first pub-

Table 1: *OOD-Speech Dataset Statistics.* ↪ denotes subsets. *WPM = Avg. Words Per Minute; WPS = Avg. Words Per Sample; OOV = Words Out of Training Vocabulary; OOG = Graphemes Out of Training Vocabulary; CS = Code-switching; NDM = Non Dominant; DM = Dominant; DS = Domain Specific; BGM = Background Music; WN = White Noise; MISC = Miscellaneous Noise; PPA = People Applauding; SV = Staged Voice; EN = Echoing Noise*

| Subset | Samples | Hours | Avg Rec. len. | WPM | WPS | Uniq. Words | OOV % | OOG % | Annot. Compl. | CS | Noise |
|---|---|---|---|---|---|---|---|---|---|---|---|
| | | | | Massively Crowdsourced Subsets | | | | | | | |
| MaCro Train | 934084 | 1129.46 | 4.35 | 133.89 | 8.42 | 207869 | 0 | 0 | - | NDM | - |
| MaCro Val | 29589 | 48.84 | 5.94 | 93.19 | 9.21 | 33055 | 0.62 | 0.2 | - | NDM | - |
| MaCro Test | 4872 | 9.84 | 7.27 | 98.35 | 11.76 | 17244 | 22.14 | 1.12 | 21.68 | NDM | - |
| | | | | Out-of-distribution Test Set | | | | | | | |
| OOD Test (Cumulative) | 2681 | 13.19 | 17.82 | 149.39 | 34.42 | 21314 | 28.03 | 5.71 | 10.60 | - | - |
| ↪ Cartoon | 399 | 2.03 | 18.30 | 115.38 | 34.55 | 569 | 65.38 | 2.64 | 13.87 | NDM | BGM, MISC |
| ↪ Online Class | 326 | 1.68 | 18.53 | 111.59 | 34.03 | 691 | 72.07 | 1.16 | 8.705 | DS | WN |
| ↪ Audiobook | 341 | 1.63 | 17.17 | 158.53 | 33.48 | 790 | 72.91 | 0.89 | 13.38 | NDM | BGM |
| ↪ Talk Show/Interview | 276 | 1.35 | 17.62 | 129.39 | 37.43 | 697 | 70.30 | 1.58 | 8.58 | NDM | BGM, WN |
| ↪ Parliament Sess. | 210 | 0.95 | 16.36 | 132.23 | 34.50 | 646 | 70.28 | 1.55 | 11.798 | NDM | WN |
| ↪ Poem Recital | 108 | 0.74 | 24.52 | 79.08 | 32.23 | 555 | 65.59 | 3.24 | 5.091 | NDM | BGM, SV, EN |
| ↪ Telemedicine | 275 | 0.73 | 9.59 | 124.69 | 20.10 | 406 | 57.64 | 0.99 | 21.254 | DS | WN |
| ↪ Beng. TV Drama | 93 | 0.68 | 26.40 | 113.55 | 50.70 | 491 | 63.34 | 0.41 | 8.561 | NDM | BGM, MISC |
| ↪ Debate | 129 | 0.64 | 17.85 | 141.96 | 42.28 | 574 | 68.47 | 3.31 | 12.64 | NDM | PPA, MISC |
| ↪ Beng. Advert. | 61 | 0.39 | 23.13 | 101.18 | 38.51 | 413 | 58.35 | 0.24 | 11.50 | DS | BGM |
| ↪ News Pres. | 62 | 0.38 | 22.03 | 124.45 | 45.53 | 470 | 63.40 | 0.21 | 7.99 | NDM | BGM |
| ↪ Ind. TV Drama | 56 | 0.35 | 22.80 | 101.51 | 38.75 | 321 | 52.65 | 0 | 8.91 | NDM | BGM |
| ↪ Slang/Profanity | 63 | 0.35 | 20.17 | 134.01 | 46.90 | 348 | 54.02 | 0.57 | 16.82 | DM | N/a |
| ↪ Movie | 48 | 0.35 | 26.04 | 103.16 | 45.33 | 377 | 57.29 | 0.27 | 14.91 | NDM | BGM, WN, MISC |
| ↪ Waz/Islamic Sermons | 41 | 0.35 | 30.34 | 117.71 | 59.51 | 362 | 54.42 | 1.10 | 5.95 | DS | WN, MISC |
| ↪ Puthi Lit. | 53 | 0.32 | 21.72 | 126.63 | 46.19 | 439 | 61.96 | 0.46 | 5.95 | NDM | N/a |
| ↪ Stage Drama | 124 | 0.28 | 8.12 | 87.05 | 10.71 | 285 | 47.37 | 0 | 16.82 | DM | PPA, BGM |

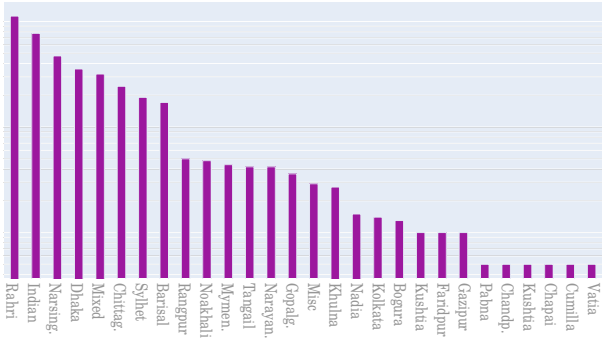

Figure 2: *Log-Distribution of different self reported dialects/accents MaCro train and MaCro validation.*

licly available OOD dataset for Bengali speech.

## 2. OOD-Speech Dataset

The OOD-Speech dataset has two different subsets delineated by whether the recorded speech was scripted or spontaneous- MaCro or **Ma**ssively **Cro**wdsourced scripted speech subset and the **O**ut-**O**f-**D**istribution (OOD) spontaneous speech subset. The MaCro set is split into three groups - train, validation and test. The OOD set is held out as a test subset of OOD-Speech. In the following subsections we describe the collection protocols, curation methods and statistics of our dataset.

### 2.1. MaCro scripted speech subset

**Collection.** The MaCro subset is collected via crowdsourcing campaigns between *Feb 2022 and Nov 2022* on the *Mozilla Common Voice* (MCV) platform [11]. Targeting national and cultural holidays in Bangladesh during the aforementioned timeframe, we conduct social network influencer marketing campaigns to draw CC0 speech contributions to MCV. The MCV platform provides a text-prompt that contributors can read out and record. The text on the platform is initially acquired by sampling from Wikipedia, which was complemented by sentences crowdsourced from volunteers. The MCV collected data is provided in the MaCro Train and Validation splits after further curation. Apart from MCV, we also created a crowdsourcing platform called *Shobdo*, through which we collect the MaCro Test split of our dataset. The data is collected through online crowdsourcing and university centric collection campaigns. The text for MaCro Test is acquired from two main sources: publicly available literary works in Bengali and public comments from YouTube, Facebook and News websites. The sentences were all manually collected and evaluated by our team of contributors before being used as prompts on the *Shobdo* platform (Suppl. Fig.11).

**Validation.** The MCV platform provides a method to upvote/downvote any recording currently on the platform. Since our crowdsourcing campaigns were focused mostly on the collection of recordings, the validated portion we obtain from MCV is quite small. To rectify this, we start by surveying 720 recordings randomly from the recordings from MCV (including samples from both the validated and invalidated portions) for qualitative analysis with a *4%* margin of error with *95%* confidence. We see that of the 720 recordings, 0.417% are unfinished recordings, 0.833% recordings are too fast to comprehend, 0.417% had loud background noise, 4.444% recordings were muffled, 0.417% were incomprehensible, 1.25% recordings had stuttering, and 0.694% contained no speech data. We use Silero [12] Voice Activity Detector (VAD) to filter out the empty recordings. We calculate the Words per minute (WPM) and manually verify and remove outliers. Therefore, we place all the human validated data in our MaCro Validation set and the rest of the data in the MaCro Training set. The MaCro test set is validated jointly with the OOD subset.

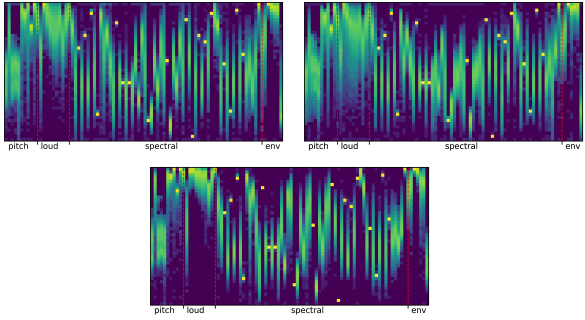

Figure 3: *Stacked log-histograms of Geneva features for MaCro Train (t. left), MaCro Test (t. right) and OpenSLR datasets(bottom).*

### 2.2. OOD spontaneous speech subset

**Sourcing.** Linguistic aspects like speed, noise, regional dialect, tonality and code switching varies between domains of speech in Bengali. We initially started by targeting 17 domains from which we wish to collect audio from, the names are provided in Tab. 1. Recordings for the *Slang/Profanity* domain were crowdsourced via a social media campaign, where participants recorded and submitted the audio anonymously. Speech for the *Telemedicine* domain was graciously contributed by the Bangladeshi govt. agency A2I. The recordings comprise anonymized phone conversations related to primary healthcare. For the other 15 domains, we created a list of 112 YouTube search queries and 22 YouTube channels as a candidate search space. Our search resulted in a list of 2768 YouTube videos which were later manually evaluated for relevance. None of the videos contained Bengali subtitles, therefore no pre-annotation was available. More details are presented in Suppl. C.

**Annotation.** A team of nine expert annotators including two medical officials for the *Telemedicine* subset, were assigned to annotate the data. Each annotator was tasked to select 5-30s windows from the YouTube sourced audio, for annotation. Annotation was performed for a period of 2 months, resulting in 13.19 hours of data.

**Validation.** We first started the validation process by assessing the Character Error Rate (CER) of a pretrained Wav2Vec2 network [13]. The top 500 words were assessed by a linguist to find the possible set of annotation errors. Following that the data was split into two segments based on a CER threshold of .4. The higher CER subset was re-annotated directly while the lower CER subset was first checked for binary correctness (yes/no) and then re-annotated if required. The authors associated with this paper, monitored and reviewed the re-annotation process. The process took a total 221h 3m by 30 expert annotators on the *Labelbox* platform.

### 2.3. Dataset Statistics

In Tab. 3 we present statistics of our MaCro and OOD subsets. For each subset we present the number of sam-ples, the combined hours of recording available, the average recording length, the average number of words per minute (WPM), average number of words in each sample (WPS), number of unique words, percentage of unique words not in training vocabulary (OOV), percentage of unique graphemes not in training vocabulary (OOG), annotation complexity, code-switching degree and noise level in decibels. We measure annotation complexity for each recording as the ratio of time required to re-annotate (as discussed in Sec. 2.2 validation) and the length of the recording. We measure the noise level as signal variance of unvoiced segments. We can see that the average recording length is significantly higher for OOD Test compared to the MaCro subsets. The WPM and WPS of OOD test is also higher than MaCro, indicating a higher rate of vocalization in spontaneous speech. Amongst the subsets of OOD Test, we see that *Audiobook* and *Online Class* subsets contain the highest number of out of training vocabulary words. Out of the 21314 unique words in the OOD Test dataset, 28.03% are out-of-distribution for the unique words in training. 5.71% of the unique graphemes are also out-of-distribution. This also indicate that there are certain phoneme clusters, corresponding to certain Bengali graphemes, that are present in the OOD Test subset but not in the MaCro train subset. The annotation complexity measure presents the average time required for annotating a second for each domain. We see that for the *Telemedicine* subset, we require the highest a We see that In Sec. 3 we further discuss these statistics and analyze the performance of trained ASR models with respect to statistics of the dataset.

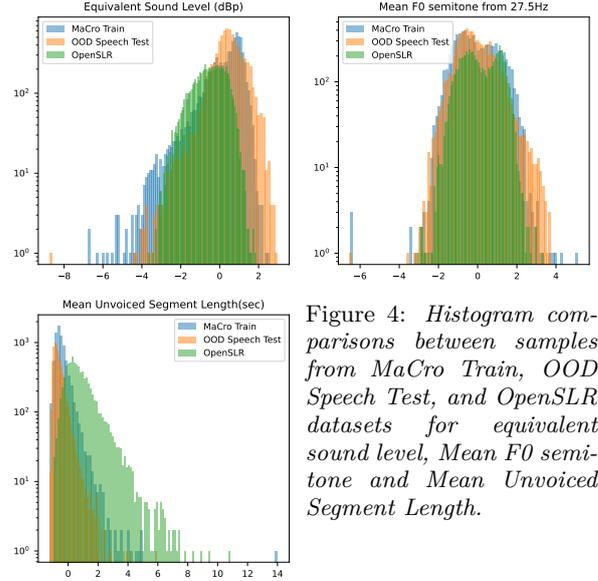

Figure 4: *Histogram comparisons between samples from MaCro Train, OOD Speech Test, and OpenSLR datasets for equivalent sound level, Mean F0 semitone and Mean Unvoiced Segment Length.*

To date, there are 5 datasets with > 100 hrs of continuous sentenceutterance level Bengali speech, that can be found in literature. Of those only 2 are publicly available. In Suppl. Tab. 3 we compare our dataset with Bengali speech Datasets with currently available Bengali Speech datasets. OOD-Speech is currently the *largest* and most *diversified publicly available* speech corpus for Bengali.

## 2.4. Speech Diversity

Diversity in speech is an important aspect of our dataset, and one of our key contributions. Our crowdsourcing campaigns for data collection on MCV were designed to ensure spatial diversity across South Asia. In Fig. 2 we present the diversity of accents present in 452 hours of recording in MaCro Train and MaCro Val, where contributors self-reported their accents. There are 28 unique self-reported accents, the most prevalent of which come from India (Rahri and Indian Bengali) followed by accents from the major districts of Bangladesh.

Some domains have high WPM, such as *Debate* with 141.96 WPM, *Telemedicine* with 124.69 WPM which portrays the reality of these domains as speakers have to finish the speech within a given amount of time, and thus they speak quickly to make the best use of time. Counter-example of low WPM can be set by the stage drama and poem recital as these represent art forms. *Telemedicine* domain has topic specific terminology and words such as names of medicine, symptoms of disease, diagnosis and treatment procedures. While *Slang/Profanity* has use of English terms, *Waz/Islamic Sermons* has frequently use of Arabic, Farsi, and Urdu words. Background music plays is common in the majority of domains, whereas *Puthi Literature* and *Slang/Profanity* domains have clear speech. In *Waz/Islamic Sermons, Debate, Stage Drama* and other domain, there may be miscellaneous noise, such as crowd participation with the speaker, noisy instrumental music, bell clanging, and so forth.

## 2.5. Feature Diversity

To assess the feature diversity of the speech in OOD-Speech, we extract Geneva speech features [7] and for 10,000 samples MaCro Train and compare it with 10,000 samples from the OpenSLR [14] speech dataset and our OOD-Speech test dataset (combining MaCro Test and OOD Test). First we use the features to compute T-Stochastic Neighbor Embeddings [6], presented in Fig. 1, Suppl. Fig. 6 and Suppl. Fig. 5. We can see in Fig. 1 the MaCro Train, MaCro Test and OOD Test subsets form separate and distinct groups, with smaller groups being formed by the subsets of OOD Test. This is a clear indication that our MaCro Test and OOD Test splits are out-of-distribution with respect to the MaCro training set. The domain wise OOD behavior is much clearer in Suppl. Fig. 5 where we show the TSNE for only the OOD Test split. If we take OpenSLR into account, we can see in Suppl. Fig. 6 that it forms a separate group from both MaCro Train and OOD-Speech Test splits. This is due to the fact that OpenSLR was collected in a controlled environment compared to our dataset which is crowdsourced from uncontrolled settings.

In Fig. 4 we present histograms for 7553 samples from MaCro Train, OpenSLR and OOD-Speech Test. The equivalent sound level for OpenSLR has a lower dynamic range compared to both MaCro Train and OOD-Speech Test. This is due to the fact that OpenSLR was acquired in a studio setting and was also post-processed. The mean unvoiced segment length for non-empty recordings, quantifies the pauses in vocalization while recording. OOD-Speech Test samples have smaller pauses compared to both MaCro Train and OpenSLR. This could be due to the presence of spontaneous speech in the OOD-Speech Test set. The Mean F0 semitone from 27.5Hz is the perceived pitch of recordings and varies based on gender and age. In Fig. 4 we can see that it is bimodal for the OpenSLR dataset but for either MaCro Train and OOD-Speech. This denotes the higher gender imbalance in our dataset compared to OpenSLR, with a larger fraction of masculine (lower pitch) recordings. Additionally, the long-term spectral average of the domain frequencies in Suppl. Fig.7 demonstrates the distinct characteristics differences of the OOD-speech domains.

Table 2: *Benchmarking pre-trained networks on the OOD-Speech test subsets.*

|  | MaCro Test | | OOD Test | | | | | | | | | |
|---|---|---|---|---|---|---|---|---|---|---|---|---|
|  | | | Audiobook | | Cartoon | | Slang/Profa. | | Talk Show/Int. | | Waz/Isl. Ser. | |
|  | WER | CER | WER | CER | WER | CER | WER | CER | WER | CER | WER | CER |
| Google | 0.41 | 0.16 | 0.55 | 0.25 | 0.77 | 0.58 | 0.79 | 0.59 | 0.60 | 0.41 | 0.74 | 0.55 |
| Wav2Vec2 | 0.38 | 0.09 | 0.52 | 0.21 | 0.74 | 0.43 | 0.76 | 0.42 | 0.58 | 0.30 | 0.70 | 0.40 |
| Conformer-CTC | 0.55 | 0.09 | 0.56 | 0.24 | 0.96 | 0.96 | 0.98 | 0.79 | 0.61 | 0.44 | 0.88 | 0.75 |
| Whisper-tiny | 0.65 | 0.51 | 0.83 | 0.62 | 0.94 | 0.74 | 1.25 | 1.51 | 0.97 | 0.90 | 0.89 | 0.74 |
| Whisper-small | 0.25 | 0.12 | 0.36 | 0.25 | 0.55 | 0.41 | 0.64 | 0.46 | 0.53 | 0.39 | 0.60 | 0.49 |
|  | Beng. Ad. | | Beng. TV | | News Pres. | | Debate | | Telemed. | | Stg. Drama | |
|  | WER | CER | WER | CER | WER | CER | WER | CER | WER | CER | WER | CER |
| Google | 0.89 | 0.81 | 0.76 | 0.61 | 0.58 | 0.33 | 0.65 | 0.44 | 0.73 | 0.61 | 0.72 | 0.54 |
| Wav2Vec2 | 0.80 | 0.53 | 0.66 | 0.38 | 0.44 | 0.18 | 0.60 | 0.32 | 0.72 | 0.43 | 0.80 | 0.48 |
| Conformer-CTC | 0.93 | 0.83 | 0.98 | 0.92 | 0.86 | 0.44 | 0.69 | 0.41 | 0.93 | 0.83 | 1.00 | 0.61 |
| Whisper-tiny | 1.0 | 0.83 | 0.91 | 0.80 | 0.79 | 0.64 | 0.79 | 0.59 | 1.09 | 0.95 | 0.90 | 0.60 |
| Whisper-small | 0.60 | 0.44 | 0.59 | 0.45 | 0.54 | 0.44 | 0.56 | 0.44 | 0.65 | 0.45 | 0.58 | 0.39 |
|  | Online Class | | Parliament | | Poem Rec. | | Movie | | Puthi Lit. | | Indian TV. | |
|  | WER | CER | WER | CER | WER | CER | WER | CER | WER | CER | WER | CER |
| Google | 0.56 | 0.34 | 0.57 | 0.34 | 0.70 | 0.46 | 0.77 | 0.66 | 0.49 | 0.21 | 0.70 | 0.55 |
| Wav2Vec2 | 0.54 | 0.23 | 0.54 | 0.30 | 0.71 | 0.37 | 0.67 | 0.40 | 0.42 | 0.12 | 0.65 | 0.35 |
| Conformer-CTC | 0.88 | 0.82 | 0.66 | 0.40 | 0.92 | 0.83 | 0.62 | 0.41 | 0.50 | 0.20 | 0.86 | 0.83 |
| Whisper-tiny | 0.95 | 0.75 | 0.93 | 0.75 | 1.55 | 1.99 | 0.93 | 0.72 | 0.73 | 0.51 | 0.89 | 0.66 |
| Whisper-small | 0.48 | 0.34 | 0.52 | 0.38 | 0.38 | 0.21 | 0.56 | 0.43 | 0.47 | 0.36 | 0.51 | 0.37 |

## 3. Benchmarking

We benchmarked the dataset using the Google speech API and Wav2Vec2 [15] model. The Wav2Vec2 model was trained on OpenSLR and fine-tuned on a subset of MaCro Validaiton [13]. For benchmarking we use Word Error Rate (WER) and Character Error Rate (CER) as metrics. The performance of the models are tabulated in Tab. 2. The models performed worse in OOD Test set across all domains when compared with MaCro Test. We also compare the performance of these models with the dataset statistics presented in Tab. 1. The models perform the worst on Bengali Advertisements, Bengali TV, Stage drama, Movie, and Cartoon, possibly indicating an effect of background music. The effect of noise can also be seen in the performance on domains with a relatively clean background, e.g., Puthi literature, Online class, Audiobook Parliament, and News presentation. The models also suffer on the Slang/Profanity domain because of the relatively higher WPM and OOV text. The models also perform badly on Poem Recital as it has one of the highest OOG percentages. This could be an indication that such networks fail to generalize on unique phonemes clusters. In general, the Wav2Vec2 model performs better than the Google API.

## 4. Conclusion

In this paper, we present OOD-Speech, the first publicly available Bengali OOD benchmarking dataset. Our dataset is also the largest publicly available Bengali ASR dataset. We provide empirical evidence on the out-of-distribution characteristics of the OOD-Speech test set with respect to its training set. This dataset can be utilized to show the robustness of an ASR model and observe the effect of domain shift on model performance. The dataset is still under development, so eradicating the currently existing issues such as gender bias, corpus sentence structure diversification will be a focus in the next version of the dataset.

## A. Extra Figures

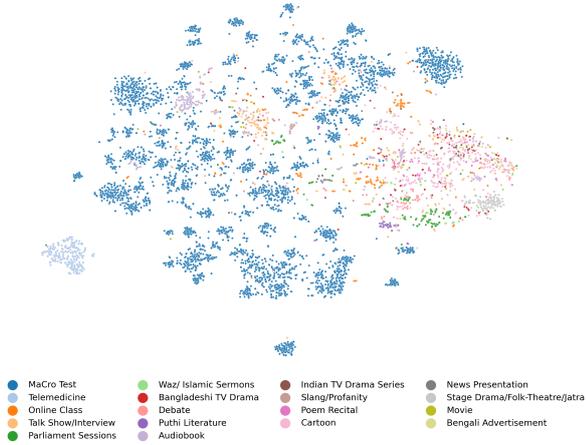

Figure 5: *Geneva feature [7] TSNE plot on the OOD test dataset with 40 perplexity and l2 metric for clustering. Separate clusters are formed for separate domains, especially from the OOD Test dataset.*

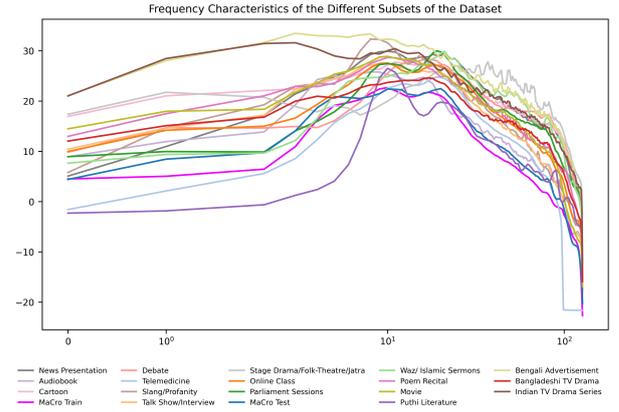

Figure 7: *Long term spectral average (LTSA) for recordings from different domains of OOD-Speech. x-axis represents the frequency domain in mel scale and y-axis presents the power spectrum in DB. Distinct envelopes of the LTSA represents the difference in frequency characteristics between domains.*

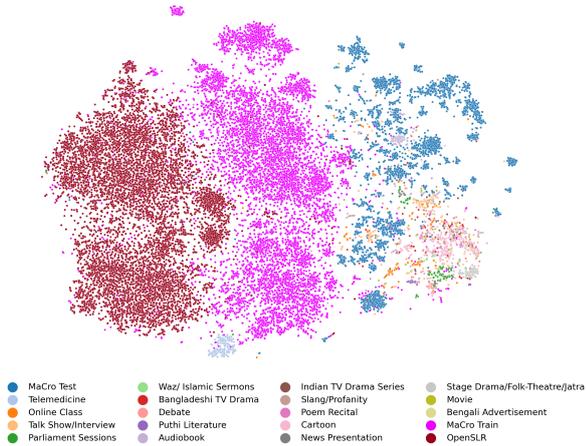

Figure 6: *Geneva feature [7] TSNE plot on the OOD test, MaCro Train, and OpenSLR dataset with 5 perplexity and l2 metric for clustering. OpenSLR forms a separated cluster from all of OOD-Speech.*

## B. Linguistic Challenges in ASR and TTS modeling of Bengali

Bengali is a language with high linguistic complexity due to its writing system, inflectional morphology, gemination, and a high number of diphthongs and triphthongs [22]. It is crucial to know these concepts to effectively design and deploy speech recognition algorithms for the Bengali language. In this section, we discuss some of the unique linguistic challenges that Bengali poses, compared to Romance or Germanic languages.

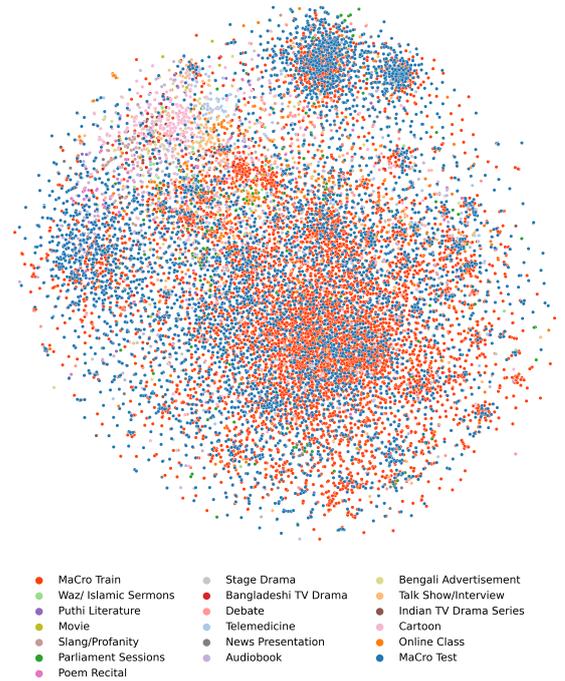

Figure 8: *t-SNE feature plot, generated utilizing the Doc2Vec [16] embeddings of transcripts from test samples and 10,000 randomly selected samples from the MaCro Train dataset. Given the broad range of topics encompassed by the MaCro Train dataset, the resulting plot exhibits significant dispersion for all subsets and no clear separations between them.*

Table 3: *Comparison of OOD-Speech with other existing Bengali speech corpus. Details on the other datasets taken from [18]*

| Year | Corpus Name | Size of Dataset | No. of Speakers | Availability |
|---|---|---|---|---|
| 2012 | IARPA-babel103b-v0.4b [8] | 215 hours | Not known | Not Publicly Available. Access per application. |
| 2014 | LDC-IL [9] | 138 hours | 240 males, 236 females | Not Publicly Available. |
| 2018 | OpenSLR [10] | 229 hours | 323 males, 182 females | CC BY-SA 3.0 US |
| 2020 | Bengali Speech Corpus from Publicly Available Audio & Text [19] | 960 hours | 268 males, 251 females | Not Publicly Available |
| 2020 | Subak.ko [20] | 241 hours | 33 males, 28 females | Not Publicly Available. |
| 2022 | Shrutilipi [21] | 441 hours | All India Radio archives | Publicly Available |
| 2022 | **OOD-Speech** | **1201.33** | **22,645+** | **CC BY-SA 3.0 US** |

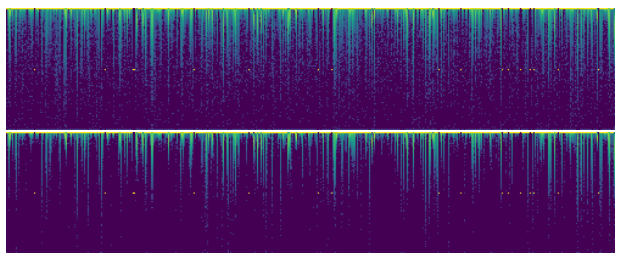

Figure 9: *Log-histograms of word recognition features for 10,000 samples from MaCro Train **(top)** and the OpenSLR Bengali ASR Dataset **(bottom)**. Features are extracted using a SpeechVGG [17] pre-trained on 1000 hours of English speech. The top histograms portray higher feature variability which is indicative of greater phoneme diversity in the MaCro Train dataset.]*

### B.1. Grapheme Diversity

One of the most unique features of the Bengali writing system is the use of orthographic syllables as graphemes. All alpha-syllabary or *Abugida* languages follow this trait. We therefore interchangeably use the terms *orthographic syllables* and *graphemes* throughout this text. For Bengali, the number of unique orthographic syllables and hence, the number of possible words are very high. Bengali contains over 1300 commonly used orthographic syllables whereas ∼ 15 million graphemes are theoretically possible [23]. Many of these occur in rare words and also due to spelling mistakes. The Bengali Unicode system also allows duplicate ways to encode the same alphabet, which requires normalization and also adds to the complexities surrounding grapheme diversity in text.

In Bengali, there are 168 commonly used grapheme roots, and of them, 80 are consonant conjuncts consisting of 2, 3 or higher consonants. A high number of consonant conjuncts leads to challenges in grapheme to phoneme mapping which depreciates the performance of TTS models, as the same conjuncts often have different pronunciations based on the context. For example, সুস্মিত is pronounced ʃuʃmit̪ whereas বিস্মিত is pronounced biʃʃit̪o, despite having the same grapheme স্মি in the middle.

### B.2. Inflection

In linguistics, *inflection* is the process of forming word variants using a template word based on grammar (e.g., variants based on tense) or sentence semantics (e.g., variants based on the gender of nouns). Like most other Indic languages, Bengali is highly inflectional. Nouns have over 100 case markers/inflection variations and verbs are found to have more than 40 unique variations. This makes the number of whitespace-tokenized 'words' occur very frequently, on the level of tens of millions compared to 188k words in English [24]. Bengali morphology also allows compound word creation which leads to a high number of Out of Vocabulary (OOV) words. This issue causes challenges in TTS and ASR modeling, especially for low-resource languages like Bengali.

### B.3. Aspiration

*Aspirated* consonants খ, ঘ, ছ, ঝ, ঠ, ঢ, থ, ধ, ফ, ভ) [$k^h$, $g^ɦ$, $c^h$, $ɟ^ɦ$, $t^h$, $d^ɦ$, $t̪^h$, $d̪^ɦ$, $p^h$, $b^ɦ$] are consonants followed by voiceless glottal fricatives [25]. Aspirated consonants pose challenges both in ASR and TTS modeling due to their subtlety in pronunciation. This, coupled with exaggerated aspiration in breathy voices, makes modeling real-life speech extremely hard [26]. Moreover, people with strong accents as well as non-native speakers, pronounce these aspirated sounds differently and hence make generalization difficult. [27].

### B.4. Gemination

*Gemination* is an instance of uttering a consonant for a longer period of time due to its consecutive use in a sentence [28]. For example, বাদ দাও [bad̪ d̪aʊ] may also be pronounced বাদ্দাও [bad̪d̪aʊ], where the latter is geminated; the difference between these two pronunciations are subtle in natural speech but are quite different in text. In Bengali, gemination may occur both in-between words and within a single word. But it is more common between two consecutive words which start and end with similar phonemes. While this linguistic feature poses challenges in determining word boundaries for ASR algorithms for many languages [29], the modeling challenges posed for Bengali are not studied in detail in the literature. Roughly, gemination in Bengali can be summarized

into the following bins:

1. Assimilation: Two sounds in a word can influence each other and reach a level of equality. জন্ম>জম্ম [jɔnmo > jɔmmo], কাঁদনা>কান্না [kãdna > kanna]

2. Progressive Assimilation: Here the previous sound of the word influences the later sound of the word. পদ্ম>পদ্দ[pɔddo > pɔddo], লগ্ন>লগ্গ[lɔgno > lɔggo]

3. Long Consonant: Sometimes to emphasize something the sounds are uttered dually in a word. পাকা>পাক্কা[paka > pakka], সকাল>সক্কাল[ʃɔkal > ʃɔkkal]

4. Deletion of Consonant Diacritic R (র): Due to the deletion of র-কার the later sound of the previous র is often uttered as a double sound of that consonant. তক>তক্ক[tɔrko > tɔkko], মারল>মাল্ল[marlo > mallo], করলাম>কল্লাম[korlam > kollam]

5. Juncture: Juncture in Bengali follows some specific rules and forms gemination, such as: বন + ওষধি = বনৌষধি [bɔn + oʃodʰi = bɔnouʃodʰi] মরু + উদ্যান = মরূদ্যান[moru + uddan = moruddan] বিপদ+জাল = বিপজ্জাল [bipɔd + jal = bipɔjjal]

## B.5. Diphthong and Triphthong

A *diphthong* is a combination formed with two vowels in a single syllable. The semi-vowel of the diphthong is placed either on the onset or the coda of the syllable. Thus diphthong is the linguistic summation of vowels plus glide. The exact number of Bengali diphthongs has varied from expert to expert [30]. According to Sukumar Sen [31], there are only two Bengali diphthongs with characters assigned to them (ঐ, ঔ). Suniti Kumar Chatterjee [32] had however pointed out 25 diphthongs while Muhammad Abdul Hai had opined that the Bengali language can have up to 31 diphthongs among which 19 are regular and 12 are irregular [33]. Bengali has a far higher number of diphthongs than English (which has 10 diphthongs).

Similarly, we have triphthongs. "Triphthong is a combination of three vowel sounds where the first vowel glides to the second which again glides to the third." as defined by Barman et al. [34]. In English, there are 5 Tripthongs whereas Bengali has 17.

Apart from these, there are many grammatical features in Bengali that make ASR and TTS modeling hard for Bengali. Brahmic Schwa deletion ambiguity [35] i.e. many Bengali words dropping their trailing vowel sounds are hard to model without a proper language model. The high number of homographs in Bengali [36] are also related to this issue making modeling more challenging. Nasality is often skipped while pronouncing by native Bengali speakers, but their written forms often explicitly include the markers of it. Also, it has been reported [37] that the phonological phrases in Bengali correspond to no plausible constituent of syntactic representation. This creates problems in modeling speech ensuring proper prosody. In Bengali speech, the phone(s) associated with glide (semivowel, glide, or semi-consonant is a sound that is phonetically similar to a vowel sound but functions as the syllable boundary), diphthongs (sound formed by the combination of two vowels) and emphasizers (modifier that serves to enhance and give additional emotional context) often sound similar. As these have different written representations, distinguishing them often requires high contextual information and this also poses challenges while modeling [38].

One unavoidable instance in Bangla sentence structure is the changing position of morpheme in the sentence. Bangla Language follows the SOV pattern of sentence structure. But in literature or daily life discourse, this pattern often alters to SVO, VSO, and OVS. [39]

Also, there are Bengali speakers who use a high number of foreign/loan words from different languages and the pronunciation and spelling of these are often not standardized. This also causes challenges while working with real-life/diversified data [40]

## C. Collection and Curation Details

### C.1. Collection protocol

*C.1.1. Data Scraping Roadmap & Prerequisites*

To construct a list of audio links for corresponding categories, the initial step in the Youtube data scraping process was to populate a list of search query and channel links. To ensure that we receive the data we need from the appropriate category and can avoid receiving irrelevant data, we carefully listed down these search query and channel links. With the 15 category data (excluding Telemedicine & Slang category), we were able to manually identify **112** different search queries as well as **22** YouTube channel links.

With the use of these resources, a script was run to scrape links from search queries and previously created channel links.
The challenges that we faced during the process are following:

- Some non-categorical links were generated as youtube algorithm works based on relevance (title, tags, descriptions and metadata of videos to determine the relevance with the search query), User engagement (Youtube determine engagement of the video by the factors of views, like, comments, watch-time), Recency (The most recent video got prioritize than the older one). And it's a fact that sometimes content creators use catchy or misinformed titles, tags, and descriptions to engage more audiences.

- An additional difficulty is that a channel may have contents of the same person, environment, audience or participants. This will prevent the entire dataset from becoming more diverse.

To overcome the challenges, we took the following measures:

- Limit the number of links generated from each search query.

- Manually validated generated links and eliminate the non-categorical links.

- For some specific categories with less content on YouTube, we have set a custom variable for scraping the links.

- It has also been maintained that we scrape from a channel that has a minimum threshold of videos and also ensure that we scrap only a few videos per channel. There were also manual checking involved to ensure the diversity in youtube channel videos (different speaker, environment, audience/participants)

Table 4: *Table of Different Domains and Characteristics Difference*

| Domain | Speed (wpm no VAD) | Code-switch (Qualitative) | OOV Level (Qual.) | Background Noise (Qual.) | Regional Dialect (Qual.) |
|---|---|---|---|---|---|
| Talk Show/Interview | 130.54 | Non Dominant | Moderate | People's racket, music, white noise | Non-Dominant, Standard Language |
| Parliament Sessions | 125.608 | Non Dominant | Moderate | White noise | Non-Dominant, Standard Language |
| News Presentation | 132.0926 | Non Dominant | High | Background music, miscellaneous noise | Non-Dominant, Standard Language |
| Stage Drama/Folk-Theatre/Jatra | 97.2539 | High | Yes | Loud instrumental music | Dominant |
| Telemedicine | 143.4904459 | Medical Domain Specific Words, Medicine Names | High | White Noise due to phone's mic | Noticeable |
| Indian TV Drama Series | 107.076 | Non Dominant | High | Loud Music | Most of the cases standard language used |
| Audio Book | 118.9696033 | Non Dominant | Moderate | Music | Sadhu language, Standard Language |
| Cartoon | 120.5870504 | Non Dominant | Moderate | Background Music, Miscellaneous Noise | Staged Voice, Dominant |
| Debate | 162.7230911 | Non Dominant | High | People Applaud, Music, Bell Ring | Standard Language |

*C.1.2. Data Sourcing*

Eventually, we were able to extract the audio data from a total **2768** youtube video links. Along with the links, we collect some metadata (Category, Search query, Video link, Video title, Length, Views, Description). Initially, we collect the data in .webm format and store the file with the video title.

The major difficulties we encountered at this step are:
- Due to webm being a lossy audio format, the audio files are highly compressed, we had to find out a audio file format which does not compress the original audio recording.
- In order to continue experimenting with the data, the stored file name presents additional difficulty related to the Unicode Normalization Issue.

To overcome the difficulties, we undertook these measures:
- The audio format of the data was transformed from webm to wav using a notebook.
- To proceed with the data we obtained, we renamed each audio file using uid as the filename, thus resolving the issues of Unicode Normalization.

*C.1.3. Data Annotation*

Following the domain-diversified data acquisition from social media platforms and gov't portals(a2i), we assembled a team of seven annotators. The team of annotators included two medical officials to annotate the Telemedicine category. The tasks were assigned to these annotators, and some rules were established to guarantee data diversity and sample quality. The rules for annotations are:

- Annotator will select a 5 to 30 second window from the audio file.
- From a single source, no more than two or three segments may well be taken.
- The phrases must form a complete sentence.
- The segments have to be selected at random.
- Time stamps should be recorded in a standard format.
- Diverse speakers or multiple speakers, audience/participant interaction, and background noise from the environment all are appreciated in the audio segments.
- The key speaker must be the primary subject of the

Table 5: *(Continuous) Table of Different Domains and Characteristics Difference*

| Domain | Speed (wpm no VAD) | Code-switch (Qualitative) | OOV Level (Qual.) | Background Noise (Qual.) | Regional Dialect (Qual.) |
|---|---|---|---|---|---|
| Bangladeshi TV Drama | 119.237934 | Non Dominant | Moderate | Background Music, Miscellaneous Noise | Dominant |
| Bengali Advertisement | 104.3555556 | Some due to product names | Moderate | Background Music, | Mostly Standard Language |
| Slang/Profanity | 145.188216 | Dominant | High | Clear voice | Dominant |
| Puthi Literature | 134.098 | Non Dominant | Moderate | Clear voice | Sadhu Language, |
| Online Class | 115.0819672 | Domain Specific Words | High | White noise | Dominant |
| Waz/ Islamic Sermons | 122.6932 | Dominant: Due to frequent use of Arabic, Farsi and Urdu words | High | People Racket, White Noise, | Dominant |
| Poem Recital | 81.1417 | Non Dominant | Moderate | Music, Staged voice, Echoing noise | Non Dominant |
| Bengali Movies/ Cinemas | 108.968386 | Non Dominant | Moderate | Music, White noise, Miscellaneous noise | Dominant |

transcription
- Regional dialects should be preserved.
- Diversity at the phonological level is appreciated in resources.

### C.1.4. Audio Splitting

After getting the response from the annotator with a category specific file with transcription and duration associated with it. We ran a python script to:

- Split the audio from the source file.
- Extract some metadata for each resource ('index','file', 'duration', 'transcription', 'status', 'path', 'audio length')
- Both the black transcription and the empty duration listed down were removed.

We confront the following difficulties since humans were involved in the process such as labeling mistakes and listing the erroneous time-frames.

To eliminate the concerns, a model and a manual validation approach were used.

- Generating error logs and manually checking the property and giving a fix if possible. Like: Some time-frames were not set as per instruction.
- Model validation was introduced to track down the human errors.

### C.2. Data Validation

#### C.2.1. Model Validation

We machine-validate the un-validate test dataset using Google ASR model and another model (wav2vec 2.0 [13]) and exclude blank recordings. We manually validated samples that had at least one full word difference between wav2vec 2.0 and ground truth. After this we finalize the machine-validated data.

#### C.2.2. Manual Validation

With the validation of Google and wav2vec 2.0 model, we were able to compile a list of recordings that had potential audio and transcription inconsistencies. We took two different process to investigate and mitigate the error that persists. These are:

1. Assign linguist to cross-check the model's prediction
2. With the help of a group of annotators using a platform named Labelbox.

**Linguist cross-validation:** We assign a linguist to cross-check the flagged recordings in order to validate the model's prediction. The linguist made his or her verdict based on the following factors:
**Diffs.** Predicted sentences from the mentioned model are compared to the annotator's transcription. Does the transcription differ substantially from the recording?
**Word Error Rate, WER.** is a measure of the percentage of incorrect words in the models predicted text as compared to the ground truth text of annotator's transcription.
**Character Error Rate, CER.** is a similar metric, but instead of words, it calculates the percentage of incorrect characters in the models predicted text.
**Word Insertion.** The number of words added to the annotator's transcription.
**Word Deletion.** The number of words removed to the annotator's transcription.
**Word Insertion and Deletion Total.** Summation of Word Insertion and Word Deletion property.

Table 6: *Domain Wise Stats*

| Sentence Domain | Total Words in the Sentences | Total Unique Words |
| --- | --- | --- |
| Audiobook | 11546 | 4774 |
| Cartoon | 13968 | 3524 |
| Debate | 5896 | 2204 |
| Telemedicine | 5632 | 1406 |
| Slang/Profanity | 2957 | 1146 |
| Talk Show/Interview | 10515 | 3650 |
| Stage Drama/Folk-Theatre/Jatra | 1334 | 667 |
| News Presentation | 2851 | 1546 |
| Online Class | 11115 | 3530 |
| Parliament Sessions | 7212 | 2678 |
| Waz/ Islamic Sermons | 2460 | 1110 |
| Poem Recital | 3435 | 1940 |
| Movie | 2183 | 1119 |
| Puthi Literature | 2454 | 1343 |
| Bengali Advertisement | 2348 | 1267 |
| Bangladeshi TV Drama | 4694 | 2102 |
| Indian TV Drama Series | 2179 | 984 |

Based on the factors, there were two flag that was assigned to categories the samples in the manual validation process:

**Incomplete Sentence.** The beginning of the audio was either before or after the time stamp. This caused a few words to be missing at the start or end of the audio file. Some also started on time but finished early, and vice versa.

**Incorrect Sentence.** There were spelling mistakes in a few of the transcriptions. There were one or more words missing from numerous lengthy audio recordings. Against the mentioned abnormalities, taken measures are:

- They were transcribed precisely as they were heard in the audio recording.
- Google and wav2vec 2.0 provided poor predictions for several samples. Thus, we classified these samples as validated.

**LabelBox re-validation & re-annotation:** Two different projects were created in the labelbox platform to perform the re-validation and re-annotation task. Data samples were distributed into these two projects based on some metrics, generated in the machine validation process. Then, a group of annotators were assigned to these projects who manually re-validate and re-annotate the data sample were necessary. Authors associated with this paper, monitor and reviewed throughout the process and evaluated the annotators performance. The entire process took 221h 3m, including 100h 4m of re-annotation work and 120h 59m of re-validation work by 20+ annotators.

### C.2.3. Audio Books

An audiobook is a version of a book that has been recorded with the narrator reading aloud from the book.

Different forms of Bengali literature like novels, poetry, science fiction have been included in this fold. The majority of audiobooks are recorded with read out text by a narrator in a home studio with a recorder and a comparatively quieter place of residence, while some are recorded in studios. Audiobooks pose different OOV issues due to the source and time of the original text. Along with linguistic diversity, audiobooks have pronunciation, speaker's tone, gender and regional dialect level variations. Some also have small to high background music. All of this pose a challenge for ASR models. The recordings have been crawled from the web.

*C.2.4. Talk Shows and Interview*

Bengali Talk shows and interviews pose unique challenges for ASR. Empirically, most popular talk shows are diplomatic and entertainment ones. In the talk show recordings where diplomates of two different parties are present, arguments and contentions take place. Hence, interference and overlap are found frequently. Also, as a wide variety of topics are discussed by people coming from a diverse background, code-switch and out of vocabulary words are pretty common in this fold. The recordings have been crawled from the web.

*C.2.5. Online Classes*

Online classes are usually less noisy than physical ones. Teacher-student interaction is less frequent as well. Students often ask questions in between the lectures during the classes that are taking place physically. In online class recordings, this overlap is rarely seen. Many OOV cases are found this these recordings as a diverse range of classroom recordings have been sampled. The recordings have been crawled from the web and also collected from academia collaborators.

*C.2.6. Parliament Sessions*

There's a precise pattern and structure of speech for this type of session. During the speech noises like interference of other speakers and the sound of applause and table thumping to commend or reproach the speaker are always heard. Also, the colloqial accent, and monotonous fast reading from the presenters are also some unique attributes of this domain. The recordings have been crawled from the web.

*C.2.7. Telemedicine*

Telemedicine is a term used to describe remote clinical services such as monitoring and diagnosis of the patient by a health official via telecommunication technologies. The telemedicine audio data may have technical glitches as a variety of telecommunication devices are used along with the weakness of cellular and internet connection. Variation in phonetic levels can be found in the audio recordings as their production, transmission, and perception may differ. During the communication normally the voices of two separate individuals are heard. Patients' voices may have an effect of illness. Noise and background interference commonly gets transmitted. Diverseness in dialect, pronunciation, accent are found in these types of recording due to variation of age, gender, regional dialect, speaking rates. Also topic specific terminology and words are used such as names of medicine, symptoms of disease, diagnosis and treatment procedures. We collected the recordings from Bangladesh Government and the whole dataest was annotated and validated by an MBBS doctor and a final year medical student.

*C.2.8. Waj/ Islamic Sermons*

Waj or Islamic sermons are public sermons delivered by an appointed Islamic personality on Islam and Qur'anic recitation. Speech of this genre is unique because of the fluctuation of the tone, pitch, and frequency of the speech. The emphasizing on different terms and on point emotion of the speech make the intended delivered message clearer to the audience. It is considered that a good melody of such sermons attracts the listeners and make the speakers sentimental which make them feel more attached to the religion. The recordings have been crawled from the web and randomly selected before annotation.

*C.2.9. Stage drama/ Folk-theatre/ Jatra*

Jatra is a type of folk drama that involves dramatic monologues, music, dance, and songs. Jatra is a well-liked folk dramatic genre in the Bengali-speaking areas of the Indian subcontinent, encompassing Bangladesh and the Indian states of West Bengal, Assam, and Tripura. The performance usually takes place on an open-air stage, with the audience surrounding it. Modern versions of Jatra feature loud music and highly stylized delivery with exaggerated gestures and orations. Music plays a crucial role in Jatra, and musicians sit on both sides of the stage playing instruments such as Dholak, pakhawaj, harmonium, karatal, khol, tabla, flute, cymbals, trumpets, behala (violin), and clarinet. The actors themselves mostly do the singing. However, the open-air setting can lead to background sounds from the surroundings, affecting the recorded audio quality. Sometimes, there may also be verbal exchanges between the spectators and the actors during the performance. The majority of the cast members have historically been male, and they also play female roles, although female actors started joining the cast in the recent decades. Jatra exhibits variations in dialects of the regional languages, which can cause difficulties in manual and automatic transcription due to variations in accents and pronunciations.

*C.2.10. Cartoon*

Bangla cartoons and Bangla dubbed cartoons usually use the same patterns of language usage. Since the primary target audience of these types of entertainment are children and pre-adolescents, the vocabulary choice and sentence patterns are always kept simple and easy so that they can understand the message and morals (if there are any) clearly. These shows use the least negative terms and avoid any kind of profanity.

*C.2.11. Slang/Profanity*

Slang or Profanity, is a socially offensive use of language, is words, phrases, and linguistic usages of an informal language. Profanity is an use of language which is deemed impolite, rude, indecent or culturally offensive. Some slang is common between the bengali speakers and some are used by specific regional peoples. It can show disrespect of something or someone or can be considered as an expression of strong feeling towards something. In our dataset we crowdsourced the slang data with recording from annotator/narrator. As a result, multiple regional languages or dialects can be found in the recordings as well as noise and background interference. There are some common words which can uniquely determine this domain. Along with linguistic diversity, slang has speaker tone, gender variations. One of the key features of this domain is the number of words is quite high compared to the other domains. The data for this domain has been crowdsourced through a social media campaign.

*C.2.12. Bangladeshi Drama*

There are three mediums of Bangladeshi TV dramas- Theatre, Television, and Radio. Language usage and delivery vary in these genres. In theatre Dramas, the dialogues are mostly standard Bangla and pronounced as perfectly as possible. Also, these dramas try to use the least foreign words, unless the drama is adapted from a literature piece. Adapted dramas use dialogues as it is in the books. The styles of television drama always change according to the trend and decide. In Bangladesh, 70's, the '80s, and 90's TV dramas would use standard Bangla. Since the 00's TV dramas took a turn linguistically by introducing colloquial Bangla in them. These dramas also use dialects and linguistics regionality. Language usage of Bangla radio dramas is similar to that of theatre dramas. Although radio dramas are usually original works of the director or the maker in charge most of these dramas still maintain the standard of the language. We have collected random snippets from these domains through online crawling.

*C.2.13. Indian TV Drama*

West Bengal Indian TV dramas, also known as "Bangla serials," are a popular form of entertainment in West Bengal. Compared to Bangladeshi TV drama, West Bengal TV dramas have been maintaining the Shantipuri standard Bangla in their dramas. Colloquial Bangla or regional dialects are rarely heard in these dramas. One of the distinguishing linguistic features of Bangla TV dramas is the use of a rich and expressive vocabulary. Bangla is a highly inflected language, which means that words can take on a variety of suffixes to indicate tense, aspect, and other grammatical features. This allows for a great deal of nuance and subtlety in the language used in Bangla TV dramas. Another notable linguistic feature of Bangla TV dramas is the use of poetic language and imagery. Bangla has a long and rich literary tradition, and this is reflected in the language used in Bangla TV dramas. Poetic devices such as alliteration, assonance, and metaphor are commonly used, giving the language a lyrical quality.

Finally, Bangla TV dramas also incorporate a range of dialects and registers, depending on the setting and characters. For example, characters from rural areas may speak in a different dialect than those from urban areas, and characters from different social classes may use different registers of language. This adds to the richness and authenticity of the language used in Bangla TV dramas.

*C.2.14. Puthi literature*

Puthi Literature is a unique genre of literature written in a mixed vocabulary drawn from Bangla, Arabic, Urdu, Persian and Hindi during the 18th and 19th centuries. Though the word puthi(or punthi) is derived from pustika or book, Puthi literature represents a particular type of writing dating between the 18th - 19th century. The poet's language was apparently based on the spoken dialect of the common Muslims of Hughli, Howrah, Kolkata, and 24-Parganas. Many Muslim poets of the period wrote in both sadhu(chaste) Bangla and Dobhasi Bangla. With a few exceptions, most Puthi literature was derivative, with poets drawing inspiration from Persian, Urdu and Hindi works as their sources. While borrowing from these works, they not only adopted the subjects but also many words, phrases and even syntaxes.

Puthi Literature used in the dataset are recorded by various narrators in a comparatively quieter place in a poetry reading manner. In the context of linguistic challenges, the recordings are diversified because of the variation in structure and patterns of Puthi Literature as well as the centuries of the creation. As a special genre the uttering style, tone of the narrator, emotional and artistic effect in these types of poetry along with age, gender, speaking rates add challenges for transcription.

*C.2.15. Bengali Advertisement*

Through Tv commercials, brands and promotions reach the audience smoothly and effectively. These commercials are made to be appealing to consumers and persuasive through their visuals and sound so they can grab the attention of the audience. TVCs usually use jingles and voice overs. Some TVCs only use a single instrumental piece and end the commercial with their catchphrase or axiom announced by the face or ambassador of the brand.

*C.2.16. Bengali Movies/ Cinemas*

Bangla cinema has a versatile usage of language. Like the Bangla Dramas, Bangla cinema can also be of two types: Bangladeshi Bangla cinema AKA Dhallywood and West Bengal Bangla cinema AKA Tollywood. Dialogues of Bangla cinemas vary according to the genre of the movies. The commercial ones have dialogues that will catch the attention of their target audience, the commoners. The Standard of these dialogues has been fluctuating over different decades. Different actors/ess deliver dialogues according to the role they are playing. Some common dialogues between the hero-heroine can be seen in most movies. Antagonist characters also have a common pattern of dialogue which includes vulgar jokes and outrageous remarks as well. But all these patterns and common dialogues are differently delivered by different actors/ess by their own style, which makes it a unique domain. The art films on the contrary are made considering the critique's review. Hence the dialogues in such cinemas are always written as uniquely as possible, often accompanied by background music and expressive oratory style.

*C.2.17. Poem Recital*

Bangla poetry is differentiated by not only writers but also by the decades the poems were written on. Changes take place within phonetic, morphological, and syntactic levels, even on the pragmatic level. Many inventions of out of vocabulary words are seen mostly on poetries. Also, when it comes to stylistics, Bangla poems often break the SOV pattern of Bangla syntax. Writing styles and usage of words and sentence patterns vary from poets to poets. Postmodern poetry style differs from the modern poetry style. In many cases one poet writes in multiple styles. These different styles leave an impact in the recitation.

Musical instruments are occasionally played in the background syncing with the rhythm of recitation. Some poetry is recited in chorus by two people of opposite genders or a group of people. The recordings have been collected by online media crawling and random sampling.

*C.2.18. News Presentation/ TV News*

There are different layers in TV news such as announcements, briefing the news items or contents, narrating the news, reports presented by the reporters, statements of the people related to the report, and voice-over. Announcements and briefing is made by the news presenter in a short span of time. The briefing is usually the headlines of the news and each headline is never more than 10-12 words long. The news narration style is different in every channel and it also varies from one presenter to another. News presenters are usually groomed and tutored on how to present and read out the news before starting their careers as news presenters. In those sessions, they are taught about different styles of news presentation which include vocal training (pitch, tone, pronunciation, voice motion, etc).

Generally, the formal version of the language is used in news presentations. There can be also regional language

*C.2.19. Debate*

When the debaters debate, they linguistically have to maintain sets of rules on how to deliver their speech. Such as the rising and falling of the tone according to the intensity of the content or sentence they are uttering. Also, in formal debates, standard and only standard Bangla is used because of the decorum of the setup. As the debaters have to establish good logic and useful information in a limited time period, even though the motion of their speech tends to be faster than usual the words are pronounced as clearly as can be.

Another noticeable fact is that the debater gets applauded by the audience and his fellow teammates by clapping or knocking on the table while he adds good points in the argument. These noises add to the complexity of ASR of this domain.

*C.2.20. Social Media Texts*

Social media platforms such as Facebook, Twitter, Instagram, LinkedIn, and others enable users to create and share content or engage in social networking. They serve as a means for people to connect with friends, family, and communities, and share information, news, images, and videos with a vast audience.

To create a dataset for analysis, texts were collected from these widely used social media platforms and subjected to grammatical refinement. However, the texts exhibit a high degree of linguistic diversity, including non-standard sentences, out-of-vocabulary words, and text modality related challenges.

The texts were spoken at a normal pace and contain instances of code-switching. The regional dialects used by the various speakers vary, but no significant background noise was introduced. This domain was scripted.

*C.2.21. Literature Texts*

Bengali Literature, written in the Bengali language, is a diverse and abundant collection of works that spans over a millennium and comprises various genres, such as poetry, fiction, drama, and non-fiction. The literature is notable for its profound humanism and exploration of themes like love, loss, nature, and the human experience. Additionally, numerous works address social and political concerns, and some are instrumental in shaping Bengali cultural identity.

The texts used for this study contain words that are out of vocabulary and a mix of standard and sadhu language. Narration was performed at a normal pace, with a few instances of code-switching and regional dialects varying among the narrators. Background noise was not significant. This domain was strictly scripted.

*C.2.22. Newspaper Text*

Bangladesh Newspapers cover a wide range of topics, including politics, economics, sports, entertainment, and more. They serve as a key platform for citizens to stay informed about local, national, and international events, and to participate in public discourse. Some of the most widely read and influential newspapers in Bangladesh include the Daily Star, Prothom Alo, and the Bangladesh Observer.

This text contains a mix of both formal and casual language, as well as transliterated words and words that are out of vocabulary (OOV). The texts were narrated at a normal pace and include a few instances of code-switching. The regional dialects vary among the different narrators. There was no significant background noise detected. This domain was also scripted.

Figure 10: *Labelbox platform re-annotation and re-validation process*

Figure 11: *Shobdo platform*